\documentclass[aps,pra,twocolumn,showpacs,superscriptaddress]{revtex4-1}

\usepackage{amsfonts,mathrsfs,amsmath,amsthm,amssymb,bbold}
\usepackage{epsfig}
\usepackage{bm}
\usepackage{tabulary}
\usepackage{color}
\usepackage{xcolor,verbatim}
\graphicspath{{./images/}}

\usepackage{color, colortbl}
\definecolor{LightCyan}{rgb}{0.88,1,1}
\definecolor{Gray}{gray}{0.95}

\usepackage{array}
\usepackage{tabulary}
\newcolumntype{K}[1]{>{\centering\arraybackslash}p{#1}}

\usepackage{courier}

\usepackage[english]{babel}
\usepackage{float}
\usepackage{graphicx}
\usepackage{bm}
\usepackage{braket}
\usepackage{xcolor}
\usepackage{amsfonts}
\usepackage{comment}
\usepackage{dirtytalk}
\usepackage{hyperref}
\usepackage{cleveref}
\usepackage{url}
\usepackage{bbold}
\usepackage{realboxes}

\usepackage{enumitem}
\usepackage{natbib}

\usepackage{float}

\newcommand{\tr}{\text{tr}}

\usepackage{listings}
\usepackage{xcolor}

\definecolor{bkgd}{RGB}{240,242,246}
\definecolor{ceruleanblue}{rgb}{0.16, 0.32, 0.75}
\definecolor{orange-red}{rgb}{1.0, 0.27, 0.0}
\definecolor{anotherblue}{RGB}{37,92,243}
\definecolor{blackblue}{RGB}{46,60,85}
\definecolor{goldyellow}{RGB}{199,146,12}
\lstdefinestyle{altstyle2}{
    backgroundcolor=\color{bkgd},
    basicstyle=\ttfamily\small\color{blackblue},
    breakatwhitespace=false,
    breaklines=true,
    captionpos=b,
    commentstyle=\color{goldyellow},
    keepspaces=true,
    keywordstyle=\color{orange-red},
    language=Python,
    numbersep=5pt,
    numberstyle=\tiny\color{ceruleanblue},
    showspaces=false,
    showstringspaces=false,
    showtabs=false,
    stringstyle=\color{anotherblue},
    tabsize=2
}
\begin{document}  
\title {\bf \texttt{tqix}: A toolbox for Quantum in \texttt{X}:\\
Quantum measurement, 
quantum tomography, 
quantum metrology, 
and others} 

\author{Le Bin Ho}
\affiliation{Ho Chi Minh City Institute of Physics, VAST, Ho Chi Minh City, Vietnam}
\affiliation{Research Institute of Electrical Communication, 
Tohoku University, Sendai, 980-8577, Japan}
\thanks{Electronic address: binho@riec.tohoku.ac.jp\\
Program's website: https://vqisinfo.wixsite.com/tqix}

\author{Kieu Quang Tuan}
\affiliation{University of Science, VNUHCM, Ho Chi Minh City, Vietnam}

\author{Hung Q. Nguyen}
\affiliation{Nano and Energy Center, VNU University of Science, Vietnam National University, 120401, Hanoi, Vietnam}

\date{\today}

\begin{abstract}
We present an open-source computer program 
written in Python language for quantum measurement
and related issues. In our program, 
quantum states and operators, 
including quantum gates, 
can be developed into a 
quantum-object function 
represented by a matrix. 
Build into the program are 
several measurement schemes, 
including von Neumann measurement 
and weak measurement. 
Various numerical simulation methods 
are used to mimic the real experiment results. 
We first provide an overview of 
the program structure and then discuss 
the numerical simulation of quantum measurement. 
We illustrate the program's performance
via quantum state tomography 
and quantum metrology. 
The program is built in a general language 
of quantum physics and thus is widely adaptable 
to various physical platforms, such as quantum optics,
ion traps, superconducting circuit devices, and others. 
It is also ideal to use in classroom guidance 
with simulation and visualization 
of various quantum systems.
\end{abstract}
%
%

\maketitle

\section{Introduction}
Quantum measurement theory is a 
fundamental concept in quantum mechanics
in which allows us to predict 
(i) the probability for obtaining measurement outcomes, 
and (ii) the post-measurement state
conditioned on the obtained outcome
~\cite{busch2018,nielsen_chuang_2010}.
Throughout quantum measurement,
the hidden quantum properties will be elucidated
to the classical world~\cite{wheeler_zurek_2014}.
It thus plays a crucial role in 
the characterization of physical systems
and immensely vital for the development 
of quantum technologies, including
quantum tomography~\cite{Paris2004}, 
quantum metrology and quantum imaging
~\cite{Giovannetti2011,RevModPhys.90.035005,Maga_a_Loaiza_2019}, 
quantum sensing~\cite{RevModPhys.89.035002}, 
quantum computing
~\cite{RevModPhys.79.135,RevModPhys.82.1,nielsen_chuang_2010}, 
quantum cryptography~\cite{pir2019advances},
and others.

On the one hand, quantum measurement and
data processing allow for reconstructing 
the quantum state of the measuring system 
via a quantum state tomography~
\cite{Paris2004,PhysRevA.64.052312}.
Besides, the prediction probability 
obtained from quantum measurement also
reveals the desired parameters 
that imprint in the measuring system
in which one can estimate those parameters
via a process called quantum metrology~
\cite{Giovannetti2011,RevModPhys.90.035005}.
On the other hand, quantum measurement
has wide-range applicability for 
establishing new quantum technologies 
such as randomized benchmarking~\cite{Helsen2019}, 
calibrating quantum operations~\cite{Frank2017}, 
and experimentally validating 
quantum computing devices~\cite{Gheorghiu2019}.

A study on quantum measurement theory 
is thus increasingly important. 
Although many physical systems 
can be carried out experimentally 
with the current technologies, 
including quantum optics, ion traps, 
superconducting circuits, NV center, 
and NMR devices, ect., 
it is still essential for developing 
an analytical and numerical tool 
for quantum measurement and data processing. 
It will be a valuable tool for studying 
and analyzing various proposed 
measurement algorithms, 
enhancing quantum tomography 
and metrology, and others.

In this work, we construct and develop 
such a toolbox for quantum measurement 
and data processing,
then apply it to quantum tomography 
and quantum metrology.
We name the program by 
\Colorbox{bkgd}{\texttt{tqix}}:
a toolbox for quantum in \texttt{X}, 
where \texttt{X}
can be the quantum measurement, 
quantum tomography,
quantum metrology, and others.
Our program serves as 
a library for creating and analyzing 
a quantum system.
Indeed, it allows for constructing 
a quantum object (states and 
operators), i.e.,
a library of standard states and 
operators are build-in 
\Colorbox{bkgd}{\texttt{tqix}}.
Furthermore, various measurement sets
\footnote{A measurement set contains
one or several positive-operator-valued measures (POVMs).}
have been constructed 
to manipulate quantum measurement, 
including Pauli, Stoke, MUB-POVM, and SIC-POVM. 
Two back-ends for simulating the measurement 
results are also built. We finally illustrate the code 
in quantum tomography and quantum metrology 
using standard data-processing tools, 
such as trace distance and fidelity.

This program is different from other existing toolboxes, 
such as Qutip, which focuses on solving the dynamics of open systems~\cite{JOHANSSON20121760,JOHANSSON20131234},
and  FEYNMAN, 
which was developed in recent years
for the simulation and analysis of quantum registers 
\cite{RADTKE2008647,RADTKE2010440,FRITZSCHE20141697} 
with $n$-qubit systems.
Here, in this work, we mainly focus on quantum measurement 
(numerical method and simulation measurement results) 
and then apply it to enhance quantum tomography 
and quantum metrology.

The rest of the paper is organized as follows: 
Section~\ref{ab} describes the program's structure. 
Section~\ref{qe} discusses quantum measurement, 
including some measurement sets and back-ends. 
Sections~\ref{qst} and \ref{qme} 
are devoted to quantum tomography 
and quantum metrology, respectively. 
In Section~\ref{lim} we discuss the limitation
of the program. 
We conclude our work in section~\ref{conc},
while Appendices are devoted to 
computational codes used in the main text.

\section{Structure of the program}
\label{ab}

\subsection{Quantum object}
In quantum physics, 
a system $\mathcal{S}$ is generally characterized by 
a preparation quantum state 
and measuring observables in the Hilbert space.
The state 
is typically represented 
by a  ket vector $|\psi\rangle$ if it is pure
or a density matrix $\rho$ if it is mixed. 
Besides, a quantum operator
associated with 
a measurable observable of $\mathcal{S}$
is described by a Hermitian matrix.
They all live in the same Hilbert space
$\mathscr{H}_\mathcal{S}$
and obey standard linear algebra.

\begin{figure} [t]
\centering
\includegraphics[width=8.6cm]{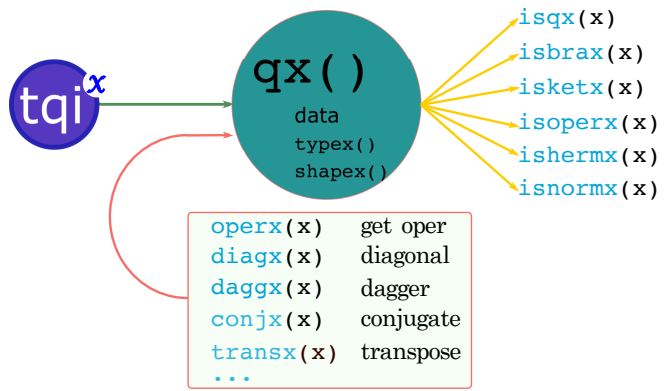}
\caption{
(Color online) Structure of a quantum object 
in \texttt{tqix}.
A quantum object can be either a bra-ket vector
or a matrice, which stands for quantum states or
operators. 
In \texttt{tqix}, a quantum object is represented by
a matrix.
We can check its physical properties 
such as type, dimension, Hermitity, and more
(see the full list in Table~\ref{t:1}).
Besides, one can also operate algebra transformations
on the quantum object such as conjugate, transpose,
and more (see the full list in Table~\ref{t:2}).
}
\label{fig1}
\end{figure}

In \Colorbox{bkgd}{\texttt{tqix}}
to represent such a quantum system $\mathcal{S}$, 
we construct a quantum object called 
\Colorbox{bkgd}{\texttt{qx}},
a matrix representation for quantum state  
and operators.
An illustration of \Colorbox{bkgd}{\texttt{qx}} 
is given in Fig.~\ref{fig1}.
A quantum object contains the data
about the given state or operator that
it represents. 
Besides providing the data, 
it also allows us to check its type and 
dimension
using 
\Colorbox{bkgd}{\texttt{typex(x)}} and 
\Colorbox{bkgd}{\texttt{shapex(x)}}, respectively. 
For example, in the following code, 
we generate a random state 
in the two-dimensional space and then check
 its type and dimension.
\begin{lstlisting}
from tqix import *
a = random(2)
print(typex(a))
print(shapex(a))
\end{lstlisting}
where we get the output as
\begin{lstlisting}
ket
(2, 1)
\end{lstlisting}
which means it is a ket (column) vector
represented by a 2 x 1 matrix.

It is easy to convert 
a given instance \texttt{x}: (integer, real, complex, tuple, array,...) 
into a quantum object using 
\Colorbox{bkgd}{\texttt{qx(x)}} command,
and its properties can be checked,
including bar, ket, oper,...
as listed in Table.~\ref{t:1}.
Furthermore, a library of 
commonly occurring operators 
is also built into
\Colorbox{bkgd}{\texttt{tqix}} 
as listed in Table.~\ref{t:2} 
and allows for operating 
on the quantum objects.
The structure of \Colorbox{bkgd}{\texttt{qx}}
is quite similar to the \Colorbox{bkgd}{\texttt{Qobj}}
class in Qutip~\cite{JOHANSSON20121760,JOHANSSON20131234}.

\begin{table}
\small
\centering
\newcolumntype{K}[1]{>{\raggedright\arraybackslash}p{#1}}
\caption {List of properties of a quantum object.
We can easy to check the physical properties 
of a quantum object by using the list below.} \label{t:1}
\setlength{\tabcolsep}{6pt}
\begin{tabular}{K{2.0cm} K{6.0cm}}
 \hline
  Method & Description \\
  \hline
 \texttt{\textcolor{cyan}{isqx}(x)} & 
  check whether \texttt{x} is a quantum object \\
  \texttt{\textcolor{cyan}{isbrax}(x)} &
  check whether \texttt{x} is a bra vector \\
  \texttt{\textcolor{cyan}{isketx}(x)} &
  check whether  \texttt{x} is a ket vector \\
 \texttt{\textcolor{cyan}{isoperx}(x)} &
  check whether \texttt{x} is an operator
  (mixed state, Hamiltonian,...) \\
   \texttt{\textcolor{cyan}{ishermx}(x)} &
  check whether \texttt{x} is a Hermit \\
  \texttt{\textcolor{cyan}{isnormx}(x)} &
  check whether \texttt{x} is normalized\\
  \hline 
\end{tabular}
\end{table}

\begin{table}
\small
\centering
\newcolumntype{K}[1]{>{\raggedright\arraybackslash}p{#1}}
\caption {List for commonly occurring operators that 
are built into \texttt{tqix}. We can easy to perform 
these familiar operators
on a quantum object.} \label{t:2}
\setlength{\tabcolsep}{6pt}
\begin{tabular}{K{2.0cm} K{6.0cm}}
 \hline
  Method & Description \\
  \hline
  \texttt{\textcolor{cyan}{operx}(x)} &
  convert a bra or ket vector into oper\\
  \texttt{\textcolor{cyan}{diagx}(x)} &
   diagonalize matrix \texttt{x}\\
   \texttt{\textcolor{cyan}{daggx}(x)} &
  get conjugate transpose of \texttt{x}: \texttt{x}$^\dagger$\\
   \texttt{\textcolor{cyan}{conjx}(x)} &
  get conjugation of \texttt{x}: \texttt{x}$^*$\\
  \texttt{\textcolor{cyan}{transx}(x)} &
  get transpose of \texttt{x}: \texttt{x}$^T$\\
  \texttt{\textcolor{cyan}{tracex}(x)} &
  get trace of \texttt{x}: (only for oper)\\
  \texttt{\textcolor{cyan}{eigenx}(x)} &
  eigenvalue and eigenstate \\
   \texttt{\textcolor{cyan}{groundx}(x)} &
   get ground state for a given Hamiltonian\\
  \texttt{\textcolor{cyan}{expx}(x)} &
  exponentiated \texttt{x}\\
  \texttt{\textcolor{cyan}{sqrtx}(x)} &
 square root of \texttt{x}\\
  \texttt{\textcolor{cyan}{l2normx}(x)} &
  get norm 2 of \texttt{x}\\
  \texttt{\textcolor{cyan}{normx}(x)} &
  get normalize of \texttt{x}\\
  \hline 
\end{tabular}
\end{table}

In subsections  \ref{sub22}
and \ref{sub23} following, 
we describe detailed quantum
states and quantum operators in 
\Colorbox{bkgd}{\texttt{tqix}}. 

\subsection{Quantum states}\label{sub22}
It is straightforward to construct quantum states 
with some standard bases and 
conventional states that are built into 
\Colorbox{bkgd}{\texttt{tqix}}. 
The list of quantum states 
can be seen in \ref{app:list_qs}.
Furthermore, to mimic a real quantum state that may contain
systematic errors or technique error, 
\Colorbox{bkgd}{\texttt{tqix}} allows us to add  
a small error to the original quantum state: 
\begin{align}
|\psi\rangle\to|\psi'\rangle
= \dfrac{1}{\mathcal N}\sum_n
(\psi_n + \delta_n)|n\rangle,
\end{align}
where $\mathcal N$ a normalization constant,
$\psi_n = \langle n|\psi\rangle$,
and $\delta_n$ is a complex random noise following
a normal distribution, e.g., 
$\delta_n = a + ib$, where 
$a, b$ are random numbers (noise).
For example,
we add small random noise into 
a GHZ state as follows:

\begin{lstlisting}
from tqix import * 
import numpy as np

psi = ghz(3)
psip = add_random_noise(psi, m = 0.0, st = 0.1)
\end{lstlisting}
Here, the random noise obeys 
a normal distribution with mean \texttt{m}
and standard deviation \texttt{st},
defaulted be zeros. 
When the noise is presented,
the quantum state will deviate from its original value.
Such noisy systems are widespread 
in various practical situations
~\cite{PhysRevLett.104.020401,Harper2020}. 

For a mixed state,
the error can be seen as a white-noise
and is given by
\begin{align}\label{e:2}
\rho\to\rho'
= (1-p)\rho + p\bm{I}/d,
\end{align}
where $p$ is a small error ($0\le p\le 1$), 
and $d$ is the dimension of the system space.
For example, one can easily add a small
white noise to an original state (e.g., GHZ state)
just with 
few lines as follows: 
\begin{lstlisting}
from tqix import *
rho = ghz(3)
rhop = add_white_noise(rho, p = 0.1)
\end{lstlisting}
where we have used \texttt{p = 0.1} 
(its default value is zero).
The function 
\Colorbox{bkgd}{\texttt{add\_white\_noise}} 
executes Eq.~\eqref{e:2} where  
its input state \texttt{rho} can be either
pure or mixed state.
An identity matrix $\bm I$ can be called from
\Colorbox{bkgd}{\texttt{tqix}} by 
\Colorbox{bkgd}{\texttt{eyex(d)}}.

We emphasize that other kinds of error
can be defined and constructed by the users
themselves, such as bit flip, phase flip, 
bit-phase flip, depolarizing, amplitude damping,
and others 
(see detailed in Chap. 8 Ref.~\cite{nielsen_chuang_2010}).

\textbf {Visualization of quantum states}. 
Phase-space representation is 
a powerful tool to visualize quantum states.
Among various ways, the visualization
using the Husimi function and Wigner function
are two common methods 
that are widely used
\cite{Schmied_2011,McConnell2015,
PhysRevA.101.022318,ahmed2020classification}. 

In general, 
a 3-dimension (3D) Husimi function representation
of a given state $\rho$ is 
$Q(\alpha) = \frac{1}{\pi}\langle \alpha|\rho
|\alpha\rangle,$
where $|\alpha\rangle$ is the coherent state,
and $\alpha = x + iy$ is a complex number.
Similarly, the Wigner function is given by
$W(\alpha) = \frac{2}{\pi}\sum_{k=0}^\infty
(-1)^k\langle \alpha, k|\rho
|\alpha, k\rangle,$
where $|\alpha, k\rangle = \bm D(\alpha)|k\rangle$
is the displaced number state, and $D(\alpha)$
is the displacement operator~
\cite{PhysRevA.48.2479},

In particular cases of spin systems,
it is more convenient to visualize 
the Husimi and Wigner functions in Bloch spheres.
The Husimi function in a Bloch sphere is 
given by 
$Q(\theta, \phi) = \frac{1}{\pi}\langle \theta,\phi|\rho
|\theta,\phi\rangle,$
where $|\theta,\phi\rangle$ is the spin coherent state,
and $\theta, \phi$ are azimuthal and polar angles, respectively. 
The Wigner function in a Bloch sphere is expressed 
as~ \cite{PhysRevA.49.4101}
\begin{align}\label{wspin}
W(\theta,\phi) = \sum_{k=0}^{2j}
\sum_{q=-k}^k \rho_{kq}\bm Y_{kq}(\theta,\phi)
\end{align}
where $Y_{kq}(\theta,\phi)$ is 
the spherical harmonic, and 
$\rho_{kq} = \sum_{m=-j}^j
\sum_{m' = -j}^j \rho_{mm'}
t_{kq}^{imm'}$ is the quantum state
represented in the spherical harmonics basis 
\cite{Schmied_2011}.
Here, $\rho_{mm'}=\langle j,m|\rho|j,m'\rangle$
is the quantum state represented in the Dicke basis 
$|j,m\rangle$, 
and $t_{kq}^{imm'}=(-1)^{j-m-q}
\langle j,m;j,-m'|k,q\rangle$
is the Clebsch-Gordan coefficient
~\cite{PhysRevA.49.4101}.

\begin{figure}[t]
\begin{center}
\includegraphics[width=8.6cm]{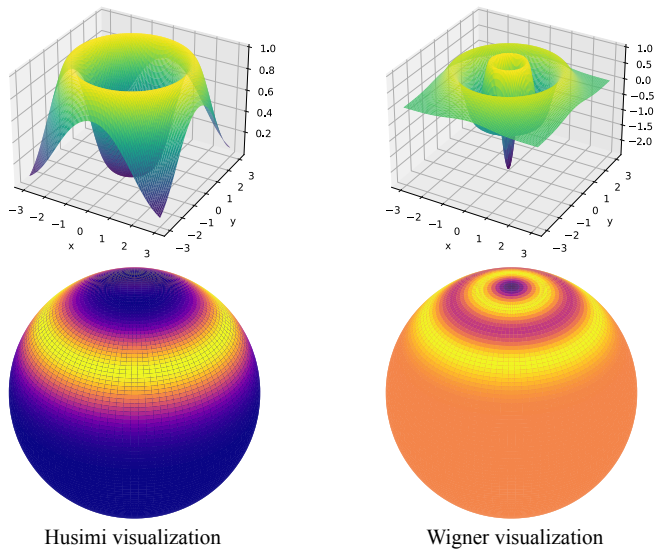}
\caption{(Color online) The Husimi and Wigner 
visualizations of a
Dicke basis $|j, m\rangle = |10,7\rangle$
in 3-dimension (upper row) 
and in the Bloch sphere (lower row).
}
\label{f:husimi}
\end{center}
\end{figure}

These visualizations are 
manipulated in \Colorbox{bkgd}{\texttt{tqix}} 
and are easy to use. 
For example, in Fig.~\ref{f:husimi}, 
we visualize the Husimi  
and Wigner functions of 
a Dicke basis $|j, m\rangle$ 
in 3D and Bloch sphere.
The visualization code is shown in 
\ref{app:husimi_visual}.

\subsection{Quantum operators}\label{sub23}

\ref{app:list_qo} represents 
some standard 
built-in quantum operators. 
In \Colorbox{bkgd}{\texttt{tqix}},
a defined operator can be either a Hamiltonian 
or an evolution operator,
represented by a matrix. 
Also, to manipulate operators' actions 
on quantum states or 
operate on multiple states, 
\Colorbox{bkgd}{\texttt{tqix}} builds
various utility mathematical functions, 
including \Colorbox{bkgd}{\texttt{dotx}} 
and \Colorbox{bkgd}{\texttt{tensorx}} 
for dot product and tensor product,
respectively. 

\subsection{Construction of quantum systems
in \texttt{tqix}}
With those standard tools presented in 
subsections  \ref{sub22} and \ref{sub23}, 
we can straightforwardly 
construct a physical system
with the given quantum state and Hamiltonian.
For example, one can construct
a two-level system state, 
e.g., $1/\sqrt{2}(|0\rangle + |1\rangle)$,
and its unitary evolution, e.g., 
$\bm U = 0.5\sigma_z - 0.25\sigma_x$,
by using the following code: 
\begin{lstlisting}
from tqix import *
from numpy import sqrt
state = 1/sqrt(2)*(obasis(2,0) + obasis(2,1))
U = 0.5*sigmaz() - 0.25*sigmax()
evolved_state = dotx(U,state) #U|psi>
\end{lstlisting}
Composite systems are also easy 
to create by using the 
\Colorbox{bkgd}{\texttt{tensorx}} 
function to generate tensor product 
states and combine Hilbert spaces.
For example, let us consider a three-spin system 
with the initial state is 
$|\uparrow\uparrow\uparrow\rangle$,
and the evolution is 
$\bm U = e^{-i\theta\sigma_z\sigma_x\sigma_x}$
where $\theta$ is a time-dependent phase.
One can use the following 
\Colorbox{bkgd}{\texttt{tqix}} code to generate
these objects:
\begin{lstlisting}
from tqix import *
from numpy import pi
theta = pi/4 #for example
up = obasis(2,0) #spin up
state = tensorx(up,up,up)
H = tensorx(sigmaz(), sigmax(), sigmax())
U = expx(-1j*theta*H)
new_state = dotx(U,state)
\end{lstlisting}


Besides, to decompose a quantum object 
(state or operator), for example,
$\rho_{\mathcal{AB}}$,
on a composite space 
$\mathscr{H}_\mathcal{A}\otimes
\mathscr{H}_\mathcal{B}$,
onto a quantum object $\rho_{\mathcal{A}}$
on $\mathscr{H}_{\mathcal{A}}$,
we can perform a partial trace
using \Colorbox{bkgd}{\texttt{ptracex}} syntax. 
It is a linear map
$\rho_\mathcal{A} = {\rm tr}_\mathcal{B}
[\rho_\mathcal{AB}]$:
$AB \to  {\rm tr}(B)A$, for any matrices $A, B$ 
on $\mathscr{H}_{\mathcal{A}}$ and 
$\mathscr{H}_{\mathcal{B}}$, respectively. 
For example, one can trace out 
the second and third subsystems
of the above Hamiltonian by using
\begin{lstlisting}
H1 = ptracex(H,[2,3])
\end{lstlisting}
Here, we keep the first subsystem.
Notable that in this version, \Colorbox{bkgd}{\texttt{ptracex}}
is only applicable for qubits systems.

\section{Quantum measurement}\label{qe}
\Colorbox{bkgd}{\texttt{tqix}}
mainly focuses on the calculation 
and simulation of quantum measurement
for quantum systems. 
We first review the quantum measurement theory 
using the POVM formalism and then describe 
how this measurement is built into 
\Colorbox{bkgd}{\texttt{tqix}}.
For simulating the measurement results, 
we also construct two available back-ends that 
can be executed in \Colorbox{bkgd}{\texttt{tqix}}. 

\subsection{General measurement}
Quantum measurement is characterized by
a set of measurement operators denoted by
$\{M_k\}$ satisfying the completeness condition
$\sum_k M^\dagger_k M_k = \bm I$, 
where $k$ is a measurement outcome. 
These measurement operators will operate
on the quantum state of the measuring system.
For a quantum system given in a general 
density state $\rho$,
the probability to obtain the outcome $k$ is
\begin{align}\label{e:pk}
p_k = {\rm tr}[M^\dagger_k M_k\ \rho ].
\end{align}
The quantum state after measurement 
will collapse to 
\begin{align}\label{e:rc}
\rho' = \dfrac{M_k \rho M^\dagger_k}{{\rm tr}[M^\dagger_k M_k\  \rho]}.
\end{align}

Furthermore, a projective measurement $\bm \Pi$ is a 
special class of the general measurement
that described by a Hermitian operator 
decomposing to 
\begin{align}\label{e:pm}
\bm\Pi = \sum_k k \Pi_k,
\end{align}
where $\Pi_k \equiv |k\rangle\langle k|$ is 
a projection operator projected onto 
the eigenstate $|k\rangle$ of $\bm \Pi$ 
with eigenvalue $k$. 
For a projective measurement, 
the probability and the post-measurement state
are calculated to be:
\begin{align}\label{e:pkpm}
p_k = {\rm tr}[\Pi_k\ \rho ], \text{ and } 
\rho' = \dfrac{\Pi_k \rho \Pi_k}{{\rm tr}[\Pi_k\  \rho]}.
\end{align}
See Ref.~\cite{nielsen_chuang_2010}
for more detailed quantum measurement.

\subsection{Positive-Operator-Valued Measurement}
In many practical cases, such as quantum tomography
and quantum metrology, 
one may not need to care about the post-measurement 
state but rather focus on the measurement probability. 
In such cases, analyzing the measurement using a 
positive-operator-valued measure (POVM) is referred.  
POVM is a mathematical description of the measurement
which is a consequence of the general measurement
\cite{nielsen_chuang_2010}.
A POVM's element, said $E_k$, 
is associated with the measurement 
$M_k$ by $M_k = W \sqrt{E_k}$,
where $W$ an arbitrary unitary operator, or
we can define
\begin{align}\label{e:povm}
E_k = M_k^\dagger M_k,
\end{align}
that is a positive operator 
satisfying the completeness
relation
 $\sum_k E_k = \bm I$. 
The probability, in this case, is given
by
\begin{align}\label{e:povm}
p_k = {\rm tr} [E_k\ \rho].
\end{align}
The POVM formalism is thus convenient for
studying the statistics of the measurement
without acknowledging 
the post-measurement state.



\textbf{Pauli measurement set}.
In quantum measurements,
one can usually combine several POVMs
as a measurement set for 
characterizing properties 
of the system to be measured.  
One common choice is the Pauli measurement set.
For one qubit, a measurement set $M$ consists of
three POVMs as
\begin{align}\label{e:m}
\bigl\{M\bigr\} = \Bigl\{|H\rangle\langle H|, |V\rangle\langle V|,
|D\rangle\langle D|, |A\rangle\langle A|,
|L\rangle\langle L|, |R\rangle\langle R|\Bigr\},
\end{align}
where
$$|H \rangle = \begin{pmatrix} 1\\0\end{pmatrix}, \ 
|V \rangle = \begin{pmatrix} 0\\1\end{pmatrix},$$
$$
|D \rangle = \dfrac{1}{\sqrt{2}}\begin{pmatrix} 1\\1\end{pmatrix}, \
|A \rangle = \dfrac{1}{\sqrt{2}}\begin{pmatrix} 1\\-1\end{pmatrix},
$$ and,
$$
|L \rangle = \dfrac{1}{\sqrt{2}}\begin{pmatrix} 1\\ i\end{pmatrix}, \
|R \rangle = \dfrac{1}{\sqrt{2}}\begin{pmatrix} 1\\-i\end{pmatrix},
$$
therein, $\{|H\rangle\langle H|, |V\rangle\langle V|\}; 
\{|D\rangle\langle D|, |A\rangle\langle A|\}$; and 
$\{|L\rangle\langle L|, |R\rangle\langle R|\Bigr\}$
are the three POVMs.
For an $n$-qubit system, the measurement
set is formed by a tensor product of elements
in $M$.
There are $6^n$ elements in total,
and thus it consumes much calculation cost
and the experimental time.  

\textbf{Stoke measurement set}.
Another practical measurement set 
is based on a light beam's polarization state, 
which was pioneering proposed by Stoke 
\cite{1851TCaPS...9..399S}.
He showed that a single qubit system 
could be determined by a set of 
four projection measurements.
These projection measurements
can be chosen arbitrarily from six elements above 
\cite{PhysRevA.64.052312,PhysRevA.66.012303}.
In \Colorbox{bkgd}{\texttt{tqix}}, we choose 
$|H\rangle\langle H|, |V\rangle\langle V|, 
|D\rangle\langle D|, \text{ and } |R\rangle\langle R|$.

Likewise the Pauli measurement set, 
all the elements of $n$-qubit system can be found
by using a tensor product of four projection measurements,
which results in $4^n$ elements.

\textbf{MUB-POVM set}.
Given two orthonormal bases $\{|e_i\rangle\}$
and $\{|f_j\rangle\}$ in a finite-dimensional Hilbert space
$\mathscr{H}_d$, they are said to be 
mutually unbiased (MUB)
if the inner product between any two elements
in each basis is a constant \cite{Schwinger570}:
\begin{align}\label{eq:ubm}
\bigl|\langle e_i|f_j\rangle\bigr|^2 = 
\dfrac{1}{d}, \forall i,j \in [1, d].
\end{align}

For a $d$-dimensional Hilbert space,  
in general, there are $d+1$ POVMs
and $d$ elements in each POVM.
In total, it needs at least 
$d^2-1$ measurements, 
which is much smaller than the Pauli case.

There are several methods for 
finding MUB-POVMs across dimension $d$,
including the Weyl group 
\cite{doi:10.1063/1.2713445},
unitary operators 
\cite{Bandyopadhyay2002},
and the Hadamard matrix method 
\cite{doi:10.1063/1.2713445}.
However, we omit writing them out here 
and encourage readers to refer 
\cite{doi:10.1142/S0219749910006502,doi:10.1063/1.2713445,
10.1007/978-3-540-24633-6_10,Bandyopadhyay2002,PhysRevA.65.032320}
if needed.
In \Colorbox{bkgd}{\texttt{tqix}},
we have constructed several MUB-POVMs for 
$d = 2, 3, 4, 5, 7$.
In the future, we also plan to develop other cases. 

\textbf{SIC-POVM set}.
In a similar manner, 
a measurement set 
$\{|h_i\rangle\}$
is called symmetric
informationally complete (SIC) 
when all the inner
products between different elements 
are equal, and their projectors
are complete 
\cite{doi:10.1063/1.1737053}:
\begin{align}\label{eq:sic}
\bigl|\langle h_i|h_j\rangle\bigr|^2 = 
\dfrac{1}{d+1}, \forall i\ne j.
\end{align}
Notable that there is only one 
POVM in a SIC-POVM set
with $d^2$ elements.

A most general way to construct 
a SIC-POVM is using the Weyl-Heisenberg 
displacement 
operators \cite{weyl_1930,doi:10.1063/1.3374022,PhysRevX.5.041006}:
\begin{align}\label{eq:weyl}
D_{j,k} = \omega_d^{jk/2}
\sum_{m=0}^{d-1}
\omega_d^{jk}
\bigl|k + m \text{ (mod d)} \big\rangle\big\langle m\bigr|,
\end{align}
where $\omega_d = \exp(2\pi i/d)$.
Then, a set of SIC-POVM elements 
can be calculated by applying the 
displacement operators on 
the fiducial vector $|\phi_d\rangle$.
The fiducial vector construction has been carried out so far
\cite{doi:10.1063/1.1896384,grassl2004sicpovms,doi:10.1063/1.1737053,Gour_2014,doi:10.1063/1.3374022,GRASSL2005151}
and is listed in~\cite{zauner}.
Such a list is also built into \Colorbox{bkgd}{\texttt{tqix}}.

\begin{figure}[t]
\begin{center}
\includegraphics[width=6.8cm]{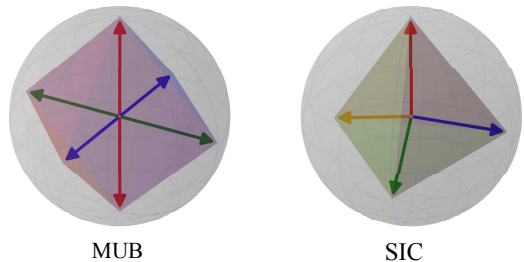}
\caption{(Color online) Bloch sphere representation 
of the MUB (a)
and SIC (b) measurement sets for one qubit.
Six measurements of MUB
correspond to vertices
of an octahedron, while 
four measurements
of SIC correspond to vertices of a tetrahedron.}
\label{f:mub_sic}
\end{center}
\end{figure}

Besides the Pauli and Stoke measurement sets,
the MUB-POVM and SIC-POVM measurement sets 
are also widely used in quantum theory
\cite{tavakoli2019mutually,PhysRevX.5.041006,Rastegin2013,Bengtsson_2010,PhysRevA.88.032312,Wootters2006,grassl2004sicpovms}. 
In Fig.~\ref{f:mub_sic}, we present a 
Bloch sphere representation of the two MUB 
and SIC measurement sets for one qubit ($d = 2$). 
Note that for one qubit, the Pauli measurement set
is the same as the MUB one.
Six elements of MUB correspond to vertices
of an octahedron, while four elements
of SIC correspond to vertices of a tetrahedron.

\subsection{Quantum measurement in \texttt{tqix}}
\label{qmt}
In \Colorbox{bkgd}{\texttt{tqix}},
one can calculate the probability 
for a given quantum system state 
and a set of observables (or POVMs)
as the following example:

\begin{lstlisting}
from tqix import *
state = ghz(1)
model = qmeas(state, [sigmax(),sigmay(),sigmaz()])
print(model.probability())
\end{lstlisting}

In this example, we calculate the expectation 
values $\langle\sigma_x\rangle, 
\langle\sigma_y\rangle,$ and $\langle\sigma_z\rangle$
of a given state $|\psi\rangle = 1/\sqrt{2} (|0\rangle + |1\rangle)$.
The outcomes are
\begin{lstlisting}
[1.0, 0.0, 0.0]
\end{lstlisting}

Furthermore, \Colorbox{bkgd}{\texttt{tqix}}
also contains the Pauli, Stoke,
MUB-POVM, and SIC-POVM measurement sets. 
One can easy to call them out by a simple syntax,
such as 
\begin{lstlisting}
model = qmeas(state, 'Pauli')
\end{lstlisting}
For other measurement sets,
one can replace \Colorbox{bkgd}{\texttt{'Pauli'}}
by \Colorbox{bkgd}{\texttt{'Stoke'}},
\Colorbox{bkgd}{\texttt{'MUB'}}, and 
\Colorbox{bkgd}{\texttt{'SIC'}}, respectively. 

\begin{figure}[t]
\begin{center}
\includegraphics[width=8.6cm]{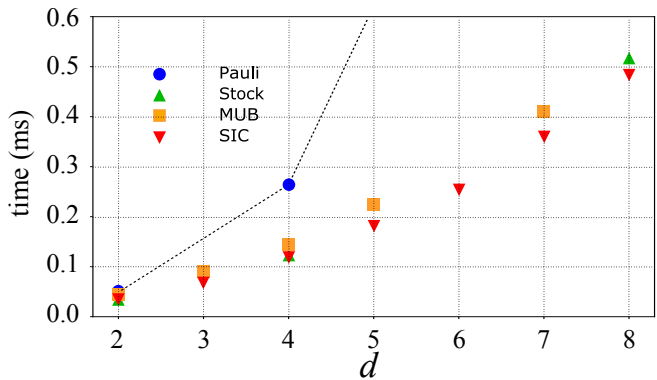}
\caption{(Color online) The calculation time of different measurement sets.
While Pauli measurement consumes much time, 
Stoke and MUB-POVM have an equivalence time, and
SIC-POVM less consumes the time.  
}
\label{f:pauli_mub_sic}
\end{center}
\end{figure}

In Fig.~\ref{f:pauli_mub_sic}, 
we compare the calculation time of 
different measurement sets for a random state in
$d$ dimension.  
Particularly, we first generate a random quantum state. 
We then measure this state via different 
measurement sets, i.e., Pauli, Stoke, MUB-POVM,
and SIC-POVM, and compare the execution time for 
each number of dimension $d$. 
See \ref{app:povm_time} for detailed code. 

\subsection{Back-end for simulation measurement results}
A back-end is defined either as a simulator or a real
device. For a given quantum object and a set of POVMs,
a back-end is executed to simulatively retrieve 
the corresponding measurement results.
In \Colorbox{bkgd}{\texttt{tqix}},
to mimic the real experiment data, 
we use several back-ends to 
simulate the results. 
A so-called Monte Carlo 
\cite{rubinstein_kroese_2017}
(\Colorbox{bkgd}{\texttt{mc}}) back-end
and
Cumulative Distribution Function 
\cite{gentle_2009}
(\Colorbox{bkgd}{\texttt{cdf}}) back-end 
have been built into \Colorbox{bkgd}{\texttt{tqix}}.
In the future, we will also construct such
a back-end from real devices. 

\textbf{mc back-end}.
Particularly, a straightforward simulation back-end used in 
\Colorbox{bkgd}{\texttt{tqix}} is 
based on the Monte Carlo simulation method,
which we named as 
\Colorbox{bkgd}{\texttt{mc}} back-end. 
Assume that a probability distribution of measurement 
is $f(x)$, generally, a function of $x$.
Without loss of generality, 
we assume that $ 0 \le f(x) \le 1, \forall x$,
(a physical probability does not exceed 
the range of [0, 1].)
For each given $f(k), k \in \{x\}$, the following
procedure is processed: \\
(\textbf{i}) generate a random number $r$ following 
a uniform distribution within [0, 1],\\
(\textbf{ii}) accept $r$ if  $r \le f(k)$,\\
(\textbf{iii}) repeat (\textbf{i}, \textbf{ii}) $N_c$ times
to get the frequency between 
the number of accepted $r$ and $N_c$,
which will distribute according
to $f(k)$.
This method has been used in quantum state tomography, for example, see Refs.~\cite{PhysRevLett.114.080403,PhysRevLett.117.010404}.

\textbf{cdf back-end}.
Another back-end is named as
\Colorbox{bkgd}{\texttt{cdf}} 
based on the cumulative distribution function 
(cdf).
For a given probability distribution of a measurement 
$f(x)$, the cdf function is
defined to be

\begin{equation}
F(y) = \int_{-\infty}^y f(x)dx.
\end{equation}

Given a uniform random variable  $r \in [0,1]$ then
$y = F^{-1}(r)$ is distributed following $f(x)$.
Like the \Colorbox{bkgd}{\texttt{mc}} back-end, 
this method has also been widely used 
in quantum state tomography~
\cite{PhysRevA.89.022122,Ho_2020,HO2019289,tuan2020direct}.

As an example, let us choose $f(x) = e^{-x}, x > 0$, 
and therefore the cumulative distribution is $F(y) = 1-e^{-y}$.
Then, for a random number $r$, the corresponding $y$ yields: 
\begin{equation}
y = -ln(1-r),
\end{equation}
which distributes according to
$f(x)$.

\begin{figure}[t]
\begin{center}
\includegraphics[width=8.6cm]{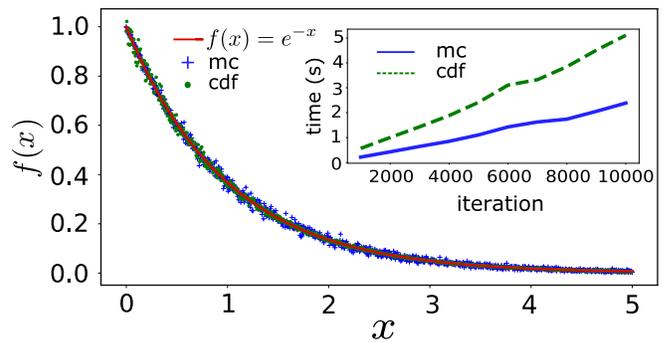}
\caption{(Color online) Plot of $f(x)$ versus 
\texttt{mc} and \texttt{cdf}
 back-ends.
Inset: the simulation time for 
\texttt{mc} and \texttt{cdf} back-ends.
While the \texttt{cdf} back-end
is more accurate than the 
\texttt{mc} back-end, 
it also consumes more time than the 
\texttt{mc} one.}
\label{f:exp}
\end{center}
\end{figure}

In Fig.~\ref{f:exp}, we plot $f(x)$ and
its simulation results via the 
\Colorbox{bkgd}{\texttt{mc}}
and \Colorbox{bkgd}{\texttt{cdf}}
back-ends. Refer to
\ref{app:test_backends} 
for detailed coding.
While the \Colorbox{bkgd}{\texttt{cdf}} back-end
is more accurate than the 
\Colorbox{bkgd}{\texttt{mc}}, 
the cost that one has to pay is that
it consumes more time than the 
\Colorbox{bkgd}{\texttt{mc}} back-end
(see the inset figure.) 
For some simulation tasks that require 
sufficient accuracy, we encourage 
the users to employ the
\Colorbox{bkgd}{\texttt{cdf}} back-end. 

\section{Quantum state tomography}
\label{qst}
\begin{figure}[t]
\begin{center}
\includegraphics[width=8.6cm]{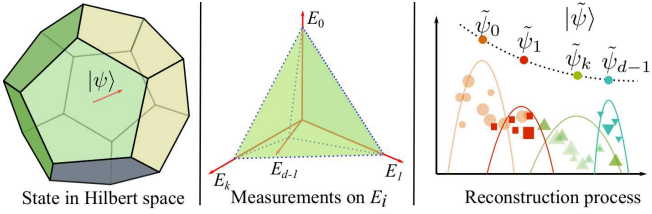}
\caption{(Color online) Quantum state tomography process:
(i) A quantum state $|\psi\rangle$ is unknown
and expressed in its Hilbert space of $d$ dimension.
(ii) The state is measured in a measurmement set given
by $E_0, E_1,\dots$.
(iii) The measurement results provide the information 
of the measuring state via an estimator, such as ML, LS,
and so forth. The reconstructed state is  
$|\tilde\psi\rangle = \sum\tilde\psi_n|n\rangle$.
}
\label{f:qst}
\end{center}
\end{figure}

In general, the quantum state tomography (QST)
is a process that reconstructing 
the information 
of unknown quantum states 
from the measurement data 
\cite{Paris2004}.
The tomography procedure 
is illustrated in Fig.~\ref{f:qst}.
A quantum state, for example $|\psi\rangle$,
is presented in its Hilbert space as given 
on the left side of the figure. 
Here, for a $d$-dimensional Hilbert space,
the state is expressed by
\begin{align}\label{eq:psi}
|\psi\rangle = \sum_{n = 0}^{d-1}
\psi_n |n\rangle,
\end{align}
where $\psi_n$ are unknown parameters
need to be estimated, and $\{|n\rangle\}$ is a 
computational basis on the Hilbert space.
The state is measured in a measurement set,
e.g., $\{E_0, E_1, \dots\}$
as illustrated in the middle of Fig.~\ref{f:qst}.
The measurement set can be chosen from
one of those described in Sec.~\ref{qe}.
Finally, the measurement results will be 
analyzed via an estimator 
such as maximum likelihood (ML)
\cite{PhysRevA.55.R1561}, 
least squares (LS) \cite{PhysRevA.64.052312}, 
neural network \cite{Torlai2018,Xin2019,PhysRevResearch.1.033157,PhysRevA.101.052316,xu2018neural,quek2018adaptive},
and others, to reproduce the state. 
The reconstructed state is denoted by
$|\tilde\psi\rangle$, where
\begin{align}\label{eq:tilde_psi}
|\tilde\psi\rangle = \sum_{n = 0}^{d-1}
\tilde\psi_n |n\rangle.
\end{align}

To evaluate the accuracy of the tomography process,
one can compare the true state $|\psi\rangle$
and the reconstructed state $|\tilde\psi\rangle$
via the various figure of merits 
such as the trace distance and the fidelity.
The trace distance between $|\psi\rangle$ and 
$|\tilde\psi\rangle$ is given by 
\begin{align}\label{eq:trace}
D(\psi,\tilde\psi) = 
\sqrt{1-\bigl|\langle\tilde\psi|\psi\rangle\bigr|^2}\;.
\end{align}
For general mixed states $\rho$
and $\tilde\rho$, the trace distance is given by
\cite{nielsen_chuang_2010}

\begin{align}\label{eq:trace_rho}
D(\rho,\tilde\rho) = \dfrac{1}{2}
{\tr}\ \bigl|\tilde\rho-\rho\bigr|\;.
\end{align}
Moreover, the fidelity is given by
\cite{nielsen_chuang_2010}
\begin{align}\label{eq:fid_rho}
F(\rho,\tilde\rho) = 
{\rm tr}\sqrt{\sqrt{\rho}\tilde\rho\sqrt{\rho}}\;.
\end{align}

\begin{figure}[t]
\begin{center}
\includegraphics[width=8.0cm]{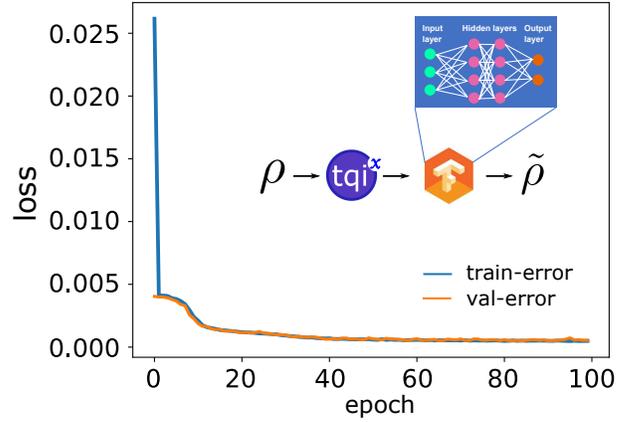}
\caption{(Color online)
Test of loss function for training data 
and validation data in \texttt{Tensorflow}.
There is no overfitting in this example 
(the training error and the validation error
are close.)
Inset: tomography scheme: 
given an unknown state $\rho$,
through \texttt{tqix}, we obtain the set of measurement 
results that can be fed into a neural network
build-in \texttt{Tensorflow}.
From the \texttt{Tensorflow}'s output,
we can reconstruct the given quantum state,
named as $\tilde\rho$.}
\label{f:t}
\end{center}
\end{figure}

Now, we give an example of using 
\Colorbox{bkgd}{\texttt{tqix}} 
in QST.
Assume that a given system is described 
by a quantum state $\rho$ that we want to
reconstruct. Using
\Colorbox{bkgd}{\texttt{tqix}} as given in 
Sec.~\ref{qmt}, 
we can get the measurement probabilities.
From those measurement data,
we reproduce the quantum state.
Here, in this example, we use a neural network scheme 
to reconstruct the quantum state, 
which is also widely used recently in the QST
\cite{Torlai2018,Xin2019,PhysRevResearch.1.033157,PhysRevA.101.052316,xu2018neural,quek2018adaptive}.
We build a neural network consists of
four layers: an input layer, two hidden layers,
and an output layer using 
\Colorbox{bkgd}{\texttt{Tensorflow}}. 
The input layer was fed by 
the set of outcome probabilities obtained from
\Colorbox{bkgd}{\texttt{tqix}}, while the output 
layer is the quantum state of being reconstructed. 
These layers were connected via a $tanh$ 
cost function.
Our scheme is illustrated in the inset Fig.
\ref{f:t}. 
In Fig.~\ref{f:t}, we also show an example of
the loss model while
training the experiment data from 
\Colorbox{bkgd}{\texttt{tqix}} with 
the Pauli measurement set.
 
 Besides, \Colorbox{bkgd}{\texttt{tqix}}
 also has been used to reconstruct 
 the quantum state using 
 the direct state measurement (DSM) method.
 For references, 
 we encourage readers to see
 Refs.~\cite{Ho_2020,HO2019289,tuan2020direct}.
 
\section{Quantum metrology}
\label{qme}
\begin{figure}[t]
\begin{center}
\includegraphics[width=8.6cm]{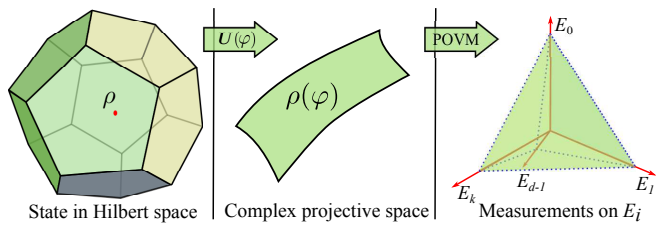}
\caption{(Color online) Quantum metrology process:
(i) A quantum system given in its density state $\rho$.
(ii) The system is exposed under an external field 
given by $\bm U(\varphi)$ where $\varphi$ 
is an unknown field parameter 
and becomes $\rho(\varphi)$. 
(iii) The state $\rho(\varphi)$ 
is measured in a 
measurement set given
by $E_0, E_1,\dots$ and 
provides the information 
on the estimated parameter $\varphi$.
}
\label{f:metro}
\end{center}
\end{figure}

Quantum metrology is a process that
unknown (single or multiples) 
parameters are estimated 
from a set of measurements~
\cite{Giovannetti2011,RevModPhys.90.035005}.
The process is illustrated in Fig.~\ref{f:metro}
for 
a single parameter estimation.
First, a system
is prepared in a general 
mixed density state $\rho$.
The state will acquire a phase $\varphi$ after 
being exposed under an external field 
represented by a unitary transformation 
$\bm U(\varphi) = e^{-i\varphi \bm H}$
and transforms to
$\rho(\varphi)= \bm U(\varphi)
\rho\bm U^\dagger(\varphi)$.
Here, $\bm H$ is a generic Hermitian operator.
The system after that will be measured 
via a POVM $\{E_k\}$, 
and the results allow for estimating 
the unknown parameter $\varphi$. 

The measurement precision (variance) 
$\Delta\varphi$
after $N$ independent measurements is defined by 
$\Delta\varphi = \sqrt{\langle 
(\varphi - \tilde\varphi)^2}\rangle$,
where $\tilde\varphi$ is the estimated value.
The minimum of $\Delta\varphi$ is 
bounded by the Cram{\'e}r-Rao bounds
\begin{align}\label{eq:C}
\Delta\varphi \ge \dfrac{1}{\sqrt{NF}}
\ge \dfrac{1}{\sqrt{NQ}},
\end{align}
where $F$ and $Q$ are the classical and 
quantum Fisher information, respectively,
which are defined by~
\cite{PhysRevLett.72.3439}
\begin{align}
F &= \sum_k\dfrac{1}{p_k(\varphi)}
\Bigl[\dfrac{\partial p_k(\varphi)}{\partial\varphi}\Bigr]^2, \label{eq:CFI} \\
Q &=2 \sum_{m,n} \dfrac{(q_m-q_n)^2}
{q_m + q_n}\bigl| \langle m|\bm H |n\rangle 
\bigr|^2. \label{eq:QFI}
\end{align}
Here, the probability of a measurement $E_k$ is given by
$p_k(\varphi) = {\rm tr} [E_k\ \rho(\varphi)],$
as in Eq.~\eqref{e:pkpm},
and the density state $\rho(\varphi)$
is given in its spectral decomposed form,
i.e., $\rho(\varphi) = \sum_m q_m |m\rangle\langle m|$,
where $q_m$ and $|m\rangle$ 
are eigenvalues and eigenvectors, respectively.

The first inequality in Eq.~\ref{eq:C} 
is classical Cram{\'e}r-Rao bound (CCRB)
while the second one is 
quantum Cram{\'e}r-Rao bound (QCRB).
In the single parameter estimation, 
the CCRB is saturated by 
the maximum likelihood estimator asymptotically 
in the number of repeated measurements 
[$N$ in Eq.~\eqref{eq:C}], 
and the QCRB is saturated by using 
an optimal measurement observable, 
which is given by the projection over 
the eigenstates of the symmetric logarithmic derivatives 
(SLD)
(see Refs.~\cite{Braunstein_1992,PhysRevLett.72.3439,
Giovannetti2011,RevModPhys.90.035005}).
Furthermore, $\Delta\varphi$ reaches 
the standard quantum limit (SQL) precision scaling
if it is proportional to $1/\sqrt{N}$,
while it reaches the Heisenberg limit (HL)
when $\Delta\varphi \propto 1/N$.

Hereafter, we provide an example 
for studying quantum metrology using
\Colorbox{bkgd}{\texttt{tqix}}.
In our example, 
we consider a cat state quantum system,
which is the superposition of spin-$j$ 
coherent states \cite{Huang2015,PhysRevA.98.012129}
\begin{align}\label{eq:cat}
|\psi\rangle = \mathcal{N}
\Bigl(|\theta,\phi\rangle
+|\pi-\theta,\phi\rangle\Bigr),
\end{align}
where $\mathcal{N}$ is the normalization constant
and $|\theta,\phi\rangle$ is a spin-$j$ coherent state
\cite{HO2019153}
\begin{align}\label{eq:spinj}
|\theta,\phi\rangle = 
\sum_{m = -j}^j
c_m
\cos^{j+m}\Bigl(\frac{\theta}{2}\Bigr)
\sin^{j-m}\Bigl(\frac{\theta}{2}\Bigr)
e^{-i(j-m)\phi}|j,m\rangle,
\end{align}
where $c_m =  \sqrt{\dfrac{(2j)!}{(j+m)!(j-m)!}}$. 
Without loss of generality, we can choose 
$\phi = 0$.
Here, $|j,m\rangle$ is the 
standard angular momentum basis
with the angular momentum quantum number
$j = 0,1/2, 1,\dots$,
and $m = -j,-j+1,\dots,j$ 
\cite{doi:10.1119/1.4898595}.

Under the transformation 
$\bm U(\varphi) = e^{-i\varphi\bm S_z}$,
the system evolves to
$|\psi(\varphi\rangle = \bm U(\varphi)
|\psi\rangle$.
 %
We then, evaluate the measurement precision by
\cite{PhysRevA.97.032116}
\begin{align}\label{eq:Delta_varphi}
\Delta\varphi =
\dfrac{
	\sqrt{\langle\bm S_y^2\rangle-\langle\bm S_y\rangle^2}
	}{
	|\partial\langle\bm S_y\rangle/\partial\varphi|
	},
\end{align}
where $\bm S_k$ ($k = x, y, z, +, -$) 
is a spin operator. 

\begin{figure}[t]
\begin{center}
\includegraphics[width=8.6cm]{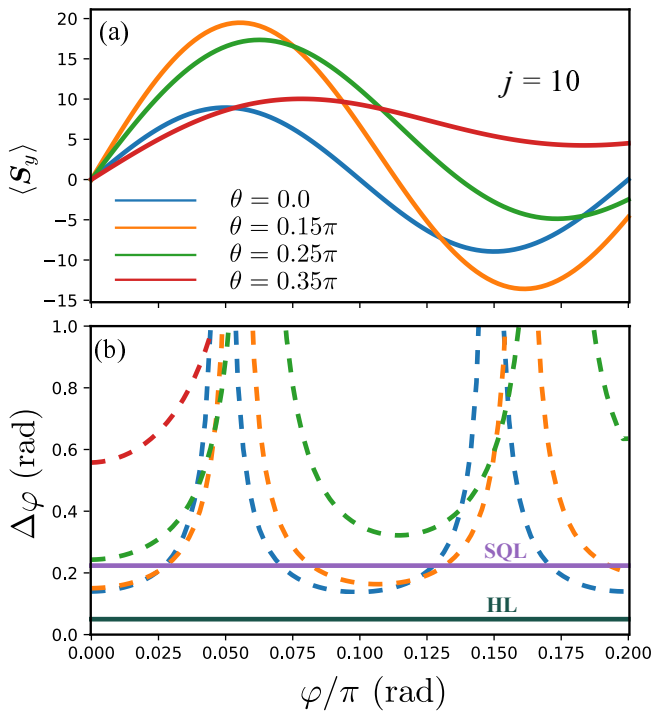}
\caption{(Color online) (a) Plot of expectation value 
$\langle\bm S_y\rangle$ as a function of 
$\varphi$. We show the results 
for several values of $\theta$ as in the figure.
Here we fix $j = 10$.
(b) Plot of the variance $\Delta\varphi$ as
a function of $\varphi$. Several results are plotted
with the same $\theta$ in (a). 
The SQL and HL are also shown in the figure.
}
\label{f:metro_ex}
\end{center}
\end{figure}

In Fig.~\ref{f:metro_ex}, we show
the expectation values $\langle\bm S_y\rangle$ (a)
and $\Delta\varphi$ (b) as functions of 
$\varphi$.
Here, we examine several values of $\theta$ as shown 
in the figure.
We can see that the variance $\Delta\varphi$
can reach the minimum at
an optimized value for 
$\varphi$.
We emphasize that $\varphi$ is a function of
the exposing time (the time that we expose the system
under the external field),
and thus there exists an optimal time 
that the variance is minimum. 
Interestingly, we can see that for 
$\theta = 0.0$ and $0.15\pi$ 
the minimum variance can beat the SQL.
This example is in agreement with
Ref.~\cite{PhysRevA.97.032116}.
The code for generating 
Fig.~\ref{f:metro_ex} is shown in
~\ref{app:metro}.
 
 \section{Limitation of the program}
\label{lim}

Within the initial version of the program, 
we cannot cover all the existing 
quantum objects (quantum states 
and operators) here. 
We limit ourselves to some basic 
and useful quantum objects as 
mentioned throughout the paper. 
Further developed versions will improve 
and complement these lacking parts.

In quantum measurement, the program limits on
some practical POVM sets, such as 
Pauli, Stoke, MUB-POVM, 
and SIC-POVM measurement sets,
which are widely used in quantum theory 
and can be carried out in various experiments.
Besides, the MUB-POVM has been 
constructed for $d$ = 2, 3, 4, 5, 7, 
where higher dimensions (to our knowledge)
do not exist yet. 

In the context of quantum tomography,
the example program is limited to 
getting meaningful quantum states such as
GHZ, W, and Dicke states.
However, we emphasize that 
the program can be used for 
estimating any given quantum states,
including non-classical states, 
such as coherent state, squeezes state,
spin-coherent state, and others. 
Here, we highlight several example calculations 
to provide the readers with an overview of the program.
Various tutorials can be found on the websites.
 
\section{Conclusion}
\label{conc}
We have presented a computer program
\Colorbox{bkgd}{\texttt{tqix}} 
by using the Python programming language
and applied to
the quantum measurement,
quantum tomography, and quantum metrology. 
In this work, we have constructed a basic
structure and some quantum features
of a quantum object which can be used
for both quantum states and quantum observables. 
There are several back-ends have been constructed
for simulation in quantum measurement. 
Thus, this program is applicable 
for a spacious range in quantum measurement, 
quantum tomography, and quantum metrology. 
We strongly encourage
those who have used 
this program to feedback 
(if any) error or incorrect to us
for  further developing the program. 

\begin{acknowledgments}
This work was supported by JSPS KAKENHI Grant Number 20F20021 and the Vietnam National University under Grant Number QG.20.17. 
\end{acknowledgments}

\begin{widetext}
\appendix

\setcounter{equation}{0}
\renewcommand{\theequation}{A\arabic{equation}}
\section{List of quantum states build-in \texttt{tqix}}
\label{app:list_qs}
In this appendix, we summary some useful 
quantum states used in the program. 
In the future, we also build 
other quantum states to the code.

\begin{table}[H]
\small
\centering
\newcolumntype{K}[1]{>{\raggedright\arraybackslash}p{#1}}
\caption {List of quantum states are built-in \texttt{tqix}} \label{t:3}
\setlength{\tabcolsep}{12pt}
\begin{tabular}{K{6.0cm} K{10cm}}
 \hline
  Name & Description \\
  \hline
 \texttt{\textcolor{cyan}{obasis}(d,k)}
  & orthogonal basis with $d$-dimension, excited at $k$,
  such as \texttt{obasis}(3,0) = $\begin{pmatrix}1\\0\\0\end{pmatrix}$, 
  \texttt{obasis}(3,1) =$\begin{pmatrix}0\\1\\0\end{pmatrix}$,...\\
\rowcolor[gray]{.9}  \texttt{\textcolor{cyan}{dbasis}(d,k)}
  & dual basis for the basis \texttt{obasis}\\
  \texttt{\textcolor{cyan}{zbasis}(j, m)}
  & Zeeman basis or Dicke basis $|j,m\rangle$ with
    spin number $j$ and quantum number $m \in [-j, j]$.\\
\rowcolor[gray]{.9}   \texttt{\textcolor{cyan}{dzbasis}(j, m)}
  & dual basis for the Zeeman basis \texttt{zbasis}.\\
  \texttt{\textcolor{cyan}{coherent}(d, alpha)}
  & generating coherent state that cut off at $d$ dimension:
  $|\alpha\rangle = e^{-\frac{|\alpha|^2}{2}}\sum_{n = 0}^d
  \frac{\alpha^n}{\sqrt{n!}}|n\rangle$, where
$\alpha$ is a complex number provided by \texttt{alpha}, 
and $\{|n\rangle\}$ is the Fock (number) basis.\\
\rowcolor[gray]{.9}  \texttt{\textcolor{cyan}{squeezed}(d, alpha, beta)}
  & generating squeezed coherent state that cut off at $d$ dimension:
  $|\alpha,\beta\rangle = 
  e^{\alpha\bm{a}^{\dagger}-
  \alpha^*\bm{a}}e^{\frac{1}{2}(\beta^*\bm{a}^2
  -{\beta\bm{a}^{\dagger}}^2)}|0\rangle$, where
  $\bm a, \bm a^\dagger$ are annihilation and 
  creation operators, respectively;
  $\alpha, \beta$ are complex numbers.
 \\
 \texttt{\textcolor{cyan}{position}(d, x)}
  & generating position state 
  which is an eigenstate of position operator $\bm{x}$, i.e.,
  $\bm{x}|x\rangle=x|x\rangle$.\\
\rowcolor[gray]{.9}  \texttt{\textcolor{cyan}{spin\_coherent}(j, theta, phi)}
  & generating spin coherent state as given in Eq.~\eqref{eq:spinj}.\\
  \texttt{\textcolor{cyan}{random}(d)}
  & random state with $d$-dimension following Haar measure.\\
  
\rowcolor[gray]{.9}  \texttt{\textcolor{cyan}{ghz}(n)}
  & generating GHZ state with $n$ qubits, i.e., 
  $\frac{1}{\sqrt{2}}(|00\cdots 0\rangle+|11\cdots 1\rangle)$.\\
  
  \texttt{\textcolor{cyan}{w}(n)}
  & generating W state with $n$ qubits, i.e.,
  $\frac{1}{\sqrt{n}}(10\cdots 0\rangle + |01\cdots 0\rangle
  + \cdots + |0\cdots 01\rangle)$.\\
  
\rowcolor[gray]{.9}  \texttt{\textcolor{cyan}{dicke}(n,k)}
  & generating Dicke state with $n$ qubits, $k$ excited qubits,
  i.e., $D(n,k)={n\choose k}^{-\frac{1}{2}}\sum_
  {x\in\{0,1\}^n}|x\rangle$. For example,
  $D(3,2) = \frac{1}{\sqrt{3}}(|011\rangle
  +|101\rangle+|110\rangle)$.\\
  \\
  \hline 
\end{tabular}
\end{table}

\setcounter{equation}{0}
\renewcommand{\theequation}{B\arabic{equation}}
\section{List of quantum operators build-in \texttt{tqix}}
\label{app:list_qo}
\begin{table}[H]
\small
\centering
\newcolumntype{K}[1]{>{\raggedright\arraybackslash}p{#1}}
\caption {List of quantum operators build-in \texttt{tqix}} \label{t:o}
\setlength{\tabcolsep}{12pt}
\begin{tabular}{K{6cm} K{10cm}}
 \hline
  Name  & Description \\
  \hline
\texttt{\textcolor{cyan}{eyex}(d)}
  & identify matrix in $d$-dimension\\
\rowcolor[gray]{.9}  \texttt{\textcolor{cyan}{soper}(s,*)}
  & spin-s operators with option *args can be $x, y, z, +, -$.
  \texttt{\textcolor{cyan}{soper}(s)} will return an array
  of spins $x, y, z$\\
 \texttt{\textcolor{cyan}{sigmax}()}
  & Pauli matrix $\sigma_x$\\
\rowcolor[gray]{.9}  \texttt{\textcolor{cyan}{sigmay}()}
  & Pauli matrix $\sigma_y$\\
  \texttt{\textcolor{cyan}{sigmaz}()}
  & Pauli matrix $\sigma_z$\\  
\rowcolor[gray]{.9}  \texttt{\textcolor{cyan}{sigmap}()}
  & $\sigma_+ $\\  
 \texttt{\textcolor{cyan}{sigmam}()}
  & $\sigma_- $\\  
\rowcolor[gray]{.9} \texttt{\textcolor{cyan}{lowering}(d)}
  & lowering (or annihilation) operator in \texttt{d} dimension\\  
  \texttt{\textcolor{cyan}{raising}(d)}
  & raising (or creation) operator in \texttt{d} dimension\\  
\rowcolor[gray]{.9}  \texttt{\textcolor{cyan}{displacement}(d, alpha)}
  & displacement operator that cut off at $d$ dimension:
  i.e., $\bm{D}(\alpha) = e^{\alpha\bm{a}^{\dagger}-
\alpha^*\bm{a}}$.
\\    
  \texttt{\textcolor{cyan}{squeezing}(d, beta)}
  & squeezing operator that cut off at $d$ dimension:
  i.e., 
$\bm{S}(\beta) = e^{\frac{1}{2}(\beta^*\bm{a}^2-{\beta\bm{a}^{\dagger}}^2)}$.\\    
  \hline 
\end{tabular}
\end{table}
\end{widetext}

\setcounter{equation}{0}
\renewcommand{\theequation}{C\arabic{equation}}
\section{Code for Husimi and Wigner visualizations (Figure \ref{f:husimi})}
\label{app:husimi_visual}

\begin{lstlisting}
from tqix import *
import numpy as np

psi = zbasis(10,7)

# 3d visualization
x = [-3, 3]
y = [-3, 3]
husimi_3d(psi, x ,y ,cmap = cmindex(1),fname ='husimi3d.eps')
wigner_3d(psi, x ,y ,cmap = cmindex(1),fname ='wigner3d.eps')

# Bloch sphere visualization 
THETA = [0, np.pi]
PHI = [0, 2* np.pi]
husimi_spin_3d(psi, THETA ,PHI ,cmap = cmindex(1),fname = 'husimi_sphere.eps')
wigner_spin_3d(psi, THETA ,PHI ,cmap = cmindex(1),fname = 'wigner_sphere.eps')
\end{lstlisting}

For \texttt{cmap}, it is ranged from \texttt{cmindex(1)} to 
\texttt{cmindex(82)} or listed in~\cite{cmap}.

\setcounter{equation}{0}
\renewcommand{\theequation}{D\arabic{equation}}
\section{Code for calculation time of POVM sets 
(Figure \ref{f:pauli_mub_sic})}
\label{app:povm_time}

\begin{lstlisting}
import time
import numpy as np
from tqix import *
import matplotlib.pyplot as plt

N = 100 # number of repeated measurement

# For Pauli and Stoke
dim_p,time_p = [],[]
time_s = [],

for n in range (1,4):
    #n: number of quibts
    dtime_p = 0.0
    dtime_s = 0.0
    
    for i in range(N):
        state = random(2**n)
        ###
        model = qmeas(state,'Pauli')
        dtime = model.mtime() #measure time
        dtime_p += dtime

        model = qmeas(state,'Stoke')
        dtime = model.mtime()
        dtime_s += dtime

    dtime_p /= float(N)
    dtime_s /= float(N)

    dim_p.append(2**n)
    time_p.append(dtime_p)
    time_s.append(dtime_s)

# For MUB
dim_mub,time_mub = [],[]
for d in (2,3,4,5,7):
    dtime_mub = 0.0
    for i in range(N):
        state = random(d)
        ###
        model = qmeas(state,'MUB')
        dtime = model.mtime()
        dtime_mub += dtime
        
    dtime_mub /= float(N)

    dim_mub.append(d)
    time_mub.append(dtime_mub)

# For SIC
dim_sic, time_sic = [],[]
for d in range(2,9):
    dtime_sic = 0.0
    for i in range(N):
        state = random(d)
        ###
        model = qmeas(state,'SIC')
        dtime = model.mtime()
        dtime_sic += dtime
        
    dtime_sic /= float(N)

    dim_sic.append(d)
    time_sic.append(dtime_sic)

# Plot figure
fig, ax1 = plt.subplots(figsize=(12,6))
ax1.plot(dim_p, time_p, marker = 'o')
ax1.plot(dim_p, time_s, marker = '^')
ax1.plot(dim_mub, time_mub, marker = 's')
ax1.plot(dim_sic, time_sic, marker = 'v')
ax1.legend(('pauli','stoke','mub','sic'))
ax1.set_xlabel('d')
ax1.set_ylabel('time (s)')
plt.savefig('time_povm.eps')
plt.show()

\end{lstlisting}

\setcounter{equation}{0}
\renewcommand{\theequation}{E\arabic{equation}}
\section{Code for back-ends (Figure \ref{f:exp})}
\label{app:test_backends}

\begin{lstlisting}
from tqix import *
import numpy as np
import matplotlib.pyplot as plt
import time

def func(n):
   fx = []
   x = np.linspace(0,5,n)
   for i in range(n):
       fx.append(np.exp(-x[i]))
   return fx

samp = 1000 #number of sample
ite  = 1000 #number iteration

fx = func(samp)
mc_sim = []
cdf_sim = []

for i in fx:
    mc_sim.append(mc(i,ite))
    
for i in fx:
    temp = []
    for j in range(ite):
        temp.append(randunit()/i)
    cdf_sim.append(cdf(temp))

x = np.linspace(0,5,samp)
fig, ax = plt.subplots(figsize=(12,6))
ax.plot(x,mc_sim,'b+')
ax.plot(x,cdf_sim,'g.')
ax.plot(x, fx, 'r')
ax.legend(("mc","cdf","fx"))
ax.set_xlabel('x')
ax.set_ylabel('exponential');
plt.savefig('fig5.eps')
plt.show()

# time check
mc_times = []
cdf_times = []
ites = []
for ite in range(1000,11000,1000):
    mc_start = time.time()
    for i in fx:
        mc_sim.append(mc(i,ite))
    mc_stop = time.time()
    delta = mc_stop - mc_start
    mc_times.append(delta)

    cdf_start = time.time()
    for i in fx:
        temp = []
        for j in range(ite):
            temp.append(randunit()/i)
        cdf_sim.append(cdf(temp))
    cdf_stop = time.time()
    delta = cdf_stop - cdf_start
    cdf_times.append(delta)

    ites.append(ite)
     
fig, ax1 = plt.subplots(figsize=(12,6))
ax1.plot(ites, mc_times, 'r')
ax1.plot(ites, cdf_times,'b--')
ax1.legend(("mc","cdf"))
ax1.set_xlabel('iteration')
ax1.set_ylabel('time (s)')
plt.savefig('inset.eps')
plt.show()
\end{lstlisting}

\section{Code for quantum metrology 
(Figure \ref{f:metro_ex})}
\label{app:metro}

\begin{lstlisting}
from tqix import *
from numpy import pi
import matplotlib.pyplot as plt

n = 20
j = n/2

# spin cat state
def cat(j,theta,phi):
    sc = spin_coherent(j,theta,phi)
    scm = spin_coherent(j,pi-theta,phi) 
    s = normx(sc + scm)
    return s

# spin observable
# j3[0] = S_x, j3[1] = S_y, j3[2] = S_z
j3 = soper(j)

def u(x):
    return np.exp(-1j*x*j3[2])

t = np.linspace(0, 0.2, 100)
theta = [0.0, 0.15*pi, 0.25*pi, 0.35*pi]
phi = 0.0

# open figures
f1 = plt.figure()
f2 = plt.figure()
ax1 = f1.add_subplot(111,aspect = 0.003)
ax2 = f2.add_subplot(111,aspect = 0.115)
for i in theta:
    r, r2, dt = [], [], []
    for k in t:
       #expectation value j3[1] and j3[1]**2
       model = qmeas(dotx(u(k*pi),cat(j,i,phi)), [j3[1],j3[1]**2]) 
       ex1 = model.probability()[0]
       ex2 = model.probability()[1]
       r.append(ex1)
       r2.append(ex2)
       dt.append(np.sqrt(np.abs(ex2-ex1**2)))

    #differential dr and delta_varphi
    dr = ndiff(t,r)
    dp = dt/np.abs(dr)
    
    #plot
    ax1.plot(t,r,'-')
    ax2.plot(t,dp,'--')

#standard quantum limit and Heisenberg limit
sql = 1./np.sqrt(float(n))*np.ones(t.shape)
hl = 1./float(n)*np.ones(t.shape)

ax2.plot(t,sql)
ax2.plot(t,hl)

plt.ylim(0,1)
plt.show()

\end{lstlisting}

\bibliographystyle{apsrev4-1}
\bibliography{refs}

\end{document}